# The Potential Impact of Digital Currencies on the Australian Economy

**Mustafa Ally**
School of Management & Enterprise
University of Southern Queensland
Queensland, Australia
Email: mustafa.ally@usq.edu.au

**Michael Gardiner**
School of Management & Enterprise
University of Southern Queensland
Queensland, Australia
Email: michael.gardiner@usq.edu.au

**Michael Lane**
School of Management & Enterprise
University of Southern Queensland
Queensland, Australia
Email: michael.lane@usq.edu.au

## Abstract

Crypto-currencies, like Bitcoins, are a relatively recent phenomena on the online Internet landscape and an emerging force in the financial sector. While not conforming to traditional institutional practices, they are gaining increasing acceptance as viable commercial currencies. With this technology presenting new opportunities, and its future largely dependent on external challenges, this conceptual paper discusses the potential impact of digital currency technology on the Australian economy. It includes (i) the payments sector, (ii) the retail sector, and (iii) the banking sector; and explores potential ways in which Australia can take advantage of digital currency technology to establish itself as a market leader in this field. The emergence of this new and potentially disruptive technology provides both opportunities as well as risks. The paper also highlights the potential impact of any tax regime that harshly penalises users of crypto-currencies. In order to support innovation and the needs of the growing Australian digital currency industry it is important to define digital currencies and examine the impact regulatory frameworks could have on the further adoption and diffusion of the technology.

## 1 Introduction

New technologies, particularly distributed peer-to-peer consensus networks and cloud-based technologies, such as the blockchain, offer the potential for valuable innovation and competition. Advances in technology have also produced rapid changes in the way Australians are managing their money (for example, PayPal, NFC, online and contactless payments, smartphone apps, etc.). The advent of digital currencies, like Bitcoin, has opened up a new range of opportunities that have the potential to support Australia's economic growth and to position Australia as a market leader in the future development of this technology (SERC 2015). Embracing digital currencies will serve to benefit the Australia economy and become a stimulus for growth and innovation. However, a number of challenges face these emerging currencies if they are to become genuine financial instruments and gain widespread adoption.

While the challenges highlighted by recent events are issues of security, usability and confidence in these currencies there have been concerns raised about the role of regulators and regulations (or lack of them) and the effect they are having in encouraging innovation and entrepreneurship in Australia (SERC 2015; Bitcoin Foundation and Bitcoin Association of Australia 2014). Tax rulings related to the use of crypto-currencies regime can have significant consequences for both consumers and businesses wanting to trade with or buy, sell and exchange Bitcoins.

Australia has the potential to become a global hub of digital currency innovation. In addition to a world-leading financial services industry, the country has a number of key market attributes conducive to growing robust digital currency industry (Bitcoin Foundation and Bitcoin Association of Australia 2014). Australia is ranked number one in the G20 countries for e-Trade Readiness (The Economist Intelligence Unit 2014). Affordable internet access, high smartphone penetration, high use of electronic payment and a well-developed regulatory framework have created a market environment that makes Australia "better placed to grow global online commerce than any other nation in the G20" (Austrade 2014).

Australia is one of the global leaders in the adoption of non-cash payments Australian Banking and Finance 2013). The Australian Clearing and Payments Association reported that the use of cash for payments in 2013 was as low as 47%, with internet and smart-phone payments making up 90% of all remote payments in 2013. This wide-spread adoption of mobile payment solutions makes Australia an ideal market in which to develop and grow a digital currency industry (Australian Payments Clearing Association 2014).

Bitcoin can play a crucial role in supporting the country's export-driven economy by reducing cross-border financial transaction fees, and by increasing the addressable market for Australian goods. From a meta-analysis analysis of the 48 submissions to the Senate Economics Review Committee and other reports and literature, this paper sets out some of the opportunities presented to the Australian payments, retail and financial sectors through exploiting the full potential of digital currencies and Bitcoins and the blockchain technology in particular in the context of the ATO rulings on taxing bitcoins and an analysis of the submissions to the Senate Economics Review Committee (SERC 2015).

## 2 Defining digital currency

The terms digital currency and virtual currency are often used interchangeably to mean the same thing. In its 2014 report on virtual currencies, the Financial Action Task Force (FATF 2014), an inter-governmental body established in 1989 by a Group of Seven (G-7) Summit in Paris, defined digital currency as a digital representation of value that can be digitally traded while functioning as a medium of exchange, unit of account and a store of value, but has no legal tender status and functions only by agreement within the community of users of the virtual currency. The European Banking Authority (2014) defined virtual currency as "a digital representation of value that is neither issued by a central bank or a public authority, nor necessarily attached to a fiat currency, but is accepted by natural or legal persons as a means of payment and can be transferred, stored or traded electronically".

### 2.1 Digital and fiat currencies

Digital currency is distinguished from fiat currency (a.k.a. 'real currency', 'real money', or 'national currency'), which is the coin and paper money of a country that is designated as its legal tender; circulates; and is customarily used and accepted as a medium of exchange in the issuing country.

Ali et al. (2014) claim that, currently, digital currencies differ from fiat money in a number of ways. A digital currency is not an IOU like fiat money. A bank holds the fiat money (liability) on behalf of a customer (asset). Also, in terms of meeting the three functions of money, namely, as a unit of account, a medium of exchange and a store of value, digital currencies fall short of their full potential. Their very limited use currently means that they are primarily seen as only a store of value.

### 2.2   Types of digital currencies

Digital currencies can further be divided into different subtypes.

#### 2.2.1   Convertible and non-convertible

A convertible digital currency has an equivalent value in fiat currency and can be exchanged back-and-forth for real currency (Linden Dollars, bitcoins, etc.). Non-convertible digital currency (closed virtual currencies with almost no link to the real economy), on the other hand, cannot be exchanged for fiat currency and is intended to be specific to a particular virtual domain, such as a massively multiplayer online role-playing game like World of Warcraft Gold which uses a non-convertible digital currency (FATF 2014). There are also virtual currencies that can be purchased directly using real currency at a specific exchange rate, but cannot be exchanged back to the original currency, for example, Facebook credits (CoinJar 2014).

#### 2.2.2   Centralized and non-centralized

All non-convertible digital currencies are centralised, as they are issued by a single administrating authority. Convertible digital currencies can be either centralised or decentralised. Decentralised digital currencies, also known as cryptocurrencies, are distributed, open-source, math-based, peer-to-peer currencies that have no central administrating authority and no central monitoring or oversight. Examples of such cryptocurrencies include: Bitcoin, Litecoin and Ripple.

## 3   Bitcoin as a digital currency

Launched in 2009, Bitcoin was the first decentralised convertible digital currency and the first cryptocurrency. Bitcoin was created as an electronic payment system that would allow two parties to transact directly with each other over the internet without needing a trusted third party intermediary. The 'distributed ledger' (also known as the 'blockchain') is used to record and verify transactions, allowing digital currency to be used as a decentralised payment system (Antonopoulos 2013).

Bitcoin as a digital currency, is gaining momentum in multiple marketplaces, bringing in benefits for both companies (lower transaction fees, instant transactions, no chargebacks, simplified payment processes) and consumers (lower or no fees to transfer value and send bitcoins globally, pseudonymous transactions, no intermediary i.e. financial institution, controlling currency) (MCDougall, 2014).

According to Winters (2014) major Australian exchanges estimate that combined they have approximately 40,000 local users. Australian adoption growth mirrors the global trend. Australia now has an estimated 192 businesses accepting Bitcoin (Winters 2014). While these numbers are low compared to the traditional banking network there is a growing uptake created by increasing user and merchant adoption. Bitcoin's market capitalisation currently sits at $6 billion USD. The Bitcoin Association of Australia estimates that the Australian share of this market capital is approximately 2%. This means that the market capitalisation for Australia is approximately $120 million.

## 4   The digital currency ecosystem

A supportive network of interconnected activities, institutions and technologies is rapidly building around virtual currencies. This developing ecosystem includes digital currency intermediaries who manage holdings and facilitate transactions. For Bitcoin users there is an ever growing range of intermediaries that provide services to users and stakeholders and, in so doing, are helping spawn new startups and entrepreneurs in this space (Figure 1).

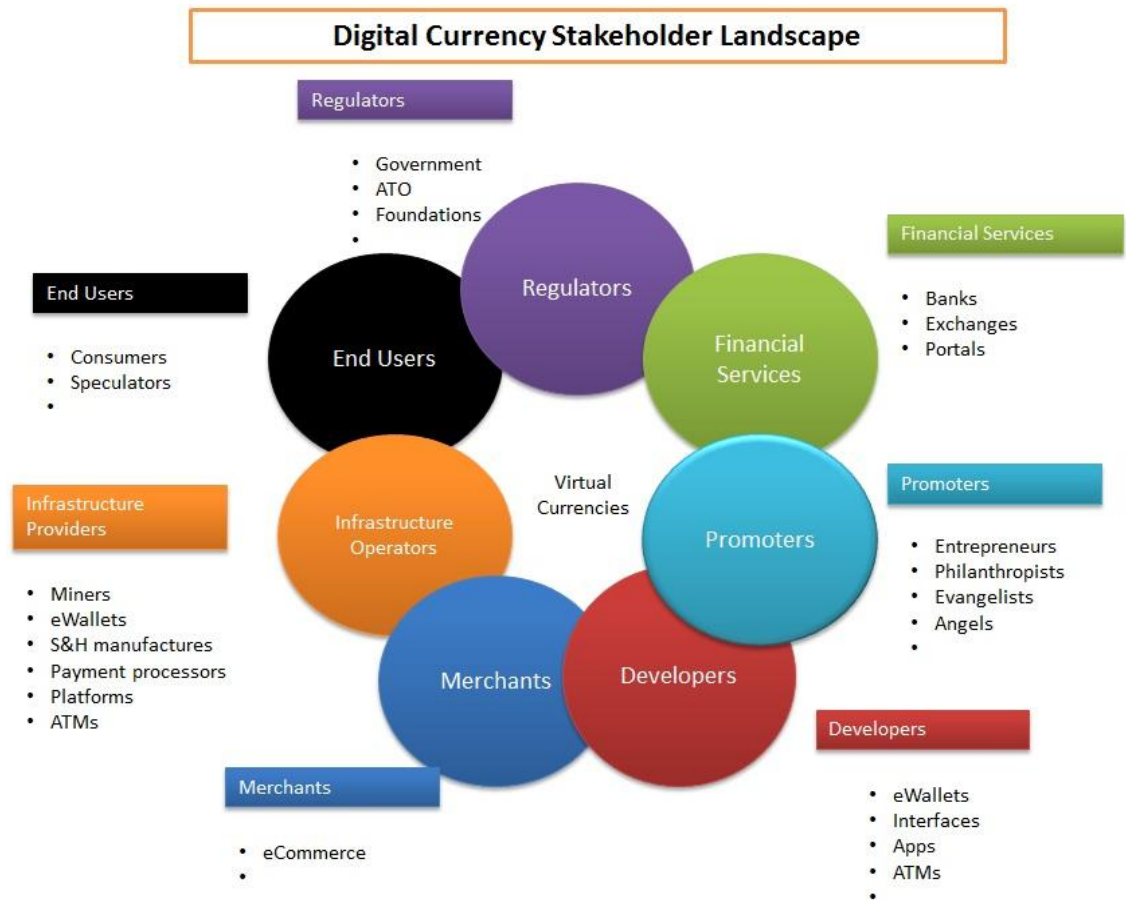

*Figure 1: Digital currency stakeholders (developed for this study)*

Value-added services can describe offerings that facilitate consumer participation in the Bitcoin ecosystem. Among the range of intermediaries are providers of bitcoin wallets, exchanges and trading platforms, merchant payment processing, Bitcoin ATMS, etc. A properly nurtured ecosystem of digital currency companies could create a range of credible small-to-medium financial providers, making the sector overall more competitive and resilient (CoinJar 2014).

## 5　Innovation opportunities

The innovation lifecycle (Figure 2) illustrates the level of technological change over time for a new innovation. A new product often leads to new processes. Beyond the blueprint and the first applications of an innovation, almost always, new innovations strive to improve, replace and complement it. Especially when the source code of such an innovation is freely available to all, this is an invitation for further development and experimentation.

The inherent technology underlying Bitcoin presents new opportunities to revolutionise how business and individuals handle modern commerce in the marketplace.

As Bitcoin begins to solidify as an innovation, we are seeing complementary processes being built on top of it, to make it more diverse (Mastercoin, Ethereum, etc.), more secure (HD wallets, multi-signature transactions, BIP38, BIP70,etc.), and more easy to use (Coinbase, BitPay, Circle, etc.).

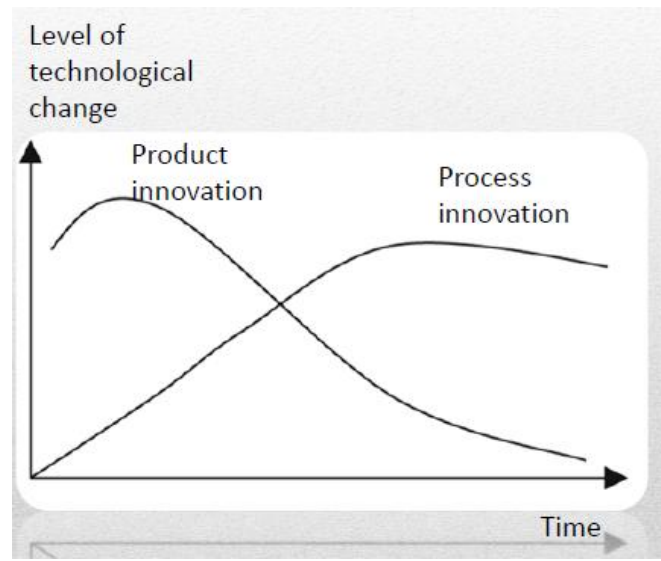

*Figure 2: Diffusion of Innovation Model (Rogers 1962)*

Four types of organizations face increased challenges if Bitcoin continues its growth pattern as a digital currency: banks, governments, payment processors and payment gateways, providing a range of opportunities for early innovators in Australia in the following three sectors:

## 5.1　Payments sector

Andreas Antonopoulos told the Senate Economics Review Committee (SERC 2015) that Bitcoin may represent a unique opportunity in two areas:

> "Firstly, bitcoin can introduce much needed competition in the retail payments industry, undercutting the expensive systems offered by credit and debit cards, while significantly improving security and privacy for consumers. Secondly, the bitcoin industry can establish Australia at the forefront of the next wave of innovation in financial services, a wave that can extend financial services to more than two billion people throughout Southeast Asia who are currently underbanked."

In the payments processing segment of the industry, merchant tools that enable online businesses to accept Bitcoin on hosted checkout pages, shopping carts and point of sale systems for brick and mortar stores can be further enhanced and developed.

Digital currencies can assist in the transmission of money offshore, and international remittances, which currently can be expensive and subject to delays in the receipt of funds, can be a growth market in Australia (APCA 2014). A case in point is that of Samoa where the transaction fees for transferring money from Australia are around 12 per cent of the transaction value. The ability to transfer value across international borders for fractions of a cent using Bitcoin technology facilitates the flow of remittances into developing countries (Bitcoin Foundation and Bitcoin Association of Australia 2014).

New business models could reduce the cost of simple transactions and increasing financial inclusion of the 2.5 billion people currently classed as the "unbanked". The unbanked are people or businesses that don't operate in the normal financial stream or system due to the high exorbitant costs banks charge or the fact that banks are not in the remote places they live in (Dostov & Shust 2014).

## 5.2   Retail sector

In addition to faster and cheaper transaction processing, the European Banking Authority (2014) identified a number of advantages attached to virtual currencies that apply particularly to the retail sector. Irreversible payments are a paradigm shift for online payment systems and the irreversible nature of Bitcoin transactions shifts the balance in favour of merchants. The certainty of payments received allows merchants to avoid having to refund transactions, particularly those based on an alleged non-fulfilment of a contract. The spawning of new types of businesses will contribute further to economic growth and the security of personal data will also foster greater trust and confidence in online ecommerce.

Micropayments, as one such example, has been long-awaited as a payment mechanism for offering very low priced content, but transaction fees have hindered deployment for anything sold for less than $1. In lieu of daily or monthly subscriptions, Bitcoin could make it possible to charge the equivalent of cents for content.

Since Bitcoin's global and decentralized digital infrastructure is not tied to country-specific currencies, travel and tourism companies could experiment with Bitcoin on multiple fronts. Tourism and vacation companies could use Bitcoin to supplement traditional payment systems, making it easier for out-of-country tourists to make reservations (PwC 2014). Airlines could accept the digital currency for domestic flight purchases in the Australia. They could also integrate QR codebased Bitcoin accounts into mobile apps, experimenting with in-flight payments, hospitality lounges or selling frequent flyer miles or points. Bitcoin could be offered as an enhancement to traditional loyalty programs such as miles or reward points (PwC 2014).

Another striking use case is online casino startups who are embracing Bitcoin, primarily for its ease of crossing national borders, absence of multiple currency conversions and lack of restrictions that banks or credit card companies may place on users for online games of chance. In offering quick, frictionless payments, digital currencies will also make it easier for retailers to experiment with new sales models, including, tipping, pay-by-use, and crowdfunding (CoinJar 2014)

## 5.3   Financial sector

Bitcoin offers unique payment options to the finance sector and has the potential to revolutionise and transform the global payments network (Branson 2014).

Bitcoin may cause financial institutions to update or add to their current technologies, adjust fee structures, add services or new layers of specialists to monitor and understand governmental regulatory issues. Blockchain technology can be leveraged to bring better efficiencies to the financial services sector with the potential of saving consumers billions of dollars annually.

Ali et al. (2014) states that digital currencies at present do not pose a significant risk to the monetary and financial stability of banks due to their size at present. The banking sector will not in the short term lose considerable business through the introduction of bitcoin, especially with the ability to obtain business through people exchanging cash for bitcoins through one of their branches. But as bitcoin grows the loss of transaction fees that they charge for businesses to use credit card and other payment methods could be considerable and worth considering, and banks in some countries may have to make their fees more comparable. Digital currencies are not expected to have an impact on central bank policy in the short term due to their size but governments and central banks are watching bitcoins progress with keen interest (Franco 2014).

Certain Australian banks have dissociated themselves from Bitcoin, closing the accounts of digital currency providers on the grounds that they posed an unacceptable level of risk, both to their business and their reputation (Southurst 2014). Exchanges, ATMs, payment processors and online merchants have the need to move money through traditional financial networks, as they bridge the present gap between the fiat and cryptocurrency systems (Southurst 2013). The banks' willingness to work with intermediaries and entrepreneurs in the digital currency world will go a long way to building a healthy and stable Bitcoin ecosystem.

## 6 Australian Taxation and Regulations

The positions of banks and other established financial institutions have for long been protected by sector specific finance regulations. Banks and credit/debit card companies are supervised by central banks, to which they also report data about the usage of various payment instruments. This is an integral part of the monetary system regulation. Digital currencies, on the other hand, by their very nature and relatively new arrival on the scence require a re-thinking of the regimes currently in place.

### 6.1 ATO Bitcoin rulings

On 20 August 2014, the Australian Taxation Office (ATO) (2014) released a suite of draft public rulings on the tax treatment of digital currencies. The ATO's rulings, which were finalised on 17 December 2014, determined that:

- Transacting with bitcoins is akin to a barter arrangement, with similar tax consequences.
- The ATO's view is that Bitcoin is neither money nor a foreign currency, and the supply of bitcoin is not a financial supply for goods and services tax (GST) purposes. Bitcoin is, however, an asset for capital gains tax (CGT) purposes.

From the Senate Economic Reference Committee Report (2015) a summary of the taxation implication of the ATO's rulings on digital currencies is as follows:

- Capital gains tax (CGT)—Those using digital currency for investment or business purposes may be subject to CGT when they dispose of digital currency, in the same way they would be for the disposal of shares or similar CGT assets; individuals who make personal use of digital currency (for example using digital currency to purchase items to buy a coffee) and where the cost of the Bitcoin was less than AUD$10,000, will have no CGT obligations.
- Goods and Services Tax (GST)—Individuals will be charged GST when they buy digital currency, as with any other property. Businesses will charge GST when they supply digital currency and be charged GST when they buy digital currency.
- Income Tax—Businesses providing an exchange service, buying and selling digital currency, or mining Bitcoin, will pay income tax on the profits. Businesses paid in Bitcoin will include the amount, valued in Australian currency, in assessable business income. Those trading digital currencies for profit, will also be required to include the profits as part of their assessable income.
- Fringe Benefits Tax (FBT)—remuneration paid in digital currency will be subject to FBT where the employee has a valid salary sacrifice arrangement, otherwise the usual salary and wage PAYG rules will apply.

### 6.2 ATO Taxation Implications

Investors see great potential to deliver good solid returns for local investment and the local community. However, Winters (2014) suggests that there are many investors who want to invest in bitcoins and their services but have been scared off due to the tax treatment of bitcoins in Australia.

As a result of the ATO decision, the Bitcoin landscape in Australia changed dramatically (Bitcoininst 2015). In December 2014, Australia's biggest cryptocurrency platform, Coinjar, relocated its headquarters to the UK in a bid to avoid GST charges on bitcoin transactions. Its chief executive, Asher Tan, said the issue had "hit the Australian bitcoin market hard", and that "several companies were forced to downsize or shut down completely" (Financial Review 2015).

According to Taxpayers Australia Limited (2014) the tax regime for digital currencies suggested by the Tax Office will:

- Negatively discriminate against businesses currently accepting Bitcoin as part of their purchasing and sales capability
- Inhibit the take-up of digital currencies across Australia, particularly amongst small businesses
- Inhibit innovation in the future development of digital currencies and associated payment systems within Australia
- Potentially lead to revenue loss through unreported or incorrectly reported transactions
- Increase compliance costs for taxpayers
- Increase administration costs within the Tax Office
- Inevitably need to be revised as current or future digital currencies become embedded into everyday payment systems

It is important for the ATO to form a logical and practical approach to the taxation of bitcoins to ensure that investment in the local Bitcoin industry is not moved offshore. This will have a significant effect on innovation in Australia in the bitcoin service industry and will leave Australia significantly behind other countries in the use of digital currencies use for the future. Removal of the GST will provide Australia's digital entrepreneurs, and the foreign businesses who want to set up in the country greater confidence and certainty in investing money and skills here (SERC 2015) and eliminate one of the important barriers to growth and innovation for digital currency in Australia.

### 6.3 Regulatory frameworks

In its Senate submission, CoinJar (2014) was of the following view:

> "One of the lessons of the music industry's battle against file-sharing is that if a digital innovation is not welcomed into a legal, regulated space, it can still thrive outside of that space. With near-instant, near-free transactions, digital currencies offer a competitive proposition that will likely thrive regardless of regulatory regime. The danger is that overly strict regulation pushes this activity into the informal economy, and offshore to territories with more accommodating regulations."

Laurel West, Editorial Director, Economist Intelligence Unit (2014) had this to say about the effect of onerous regulations: "It's clear that technology alone is not enough to allow e-trade to reach its full potential. Customs regimes across the globe are still aligned with the needs of big businesses and hampering SMEs. E-trade is a ripe opportunity for SMEs to compete with multinationals. They can be a key driver in its growth, but bureaucracy could be their biggest barrier."

The Australian Reserve Bank is the principal regulator of the payments system. Digital currencies are not currently regulated by the Bank or subject to regulatory oversight. The Reserve Bank of Australia (2014) sees that there are currently no regulatory factors on the part of the Bank that might impede growth of the digital currency industry and that this factor of regulation may well be a factor contributing to the adoption of bitcoin by some users.

The SERC (2015) examined the unique challenges that digital currencies have created for regulators, including how to maintain the integrity of the financial system while creating a regulatory environment that encourages innovation.

The committee was of the view that any regulatory framework should balance the need to mitigate risks facing consumers and the broader financial system, while still encouraging innovation and growth in the industry by keeping the barriers to entry low. As the digital currency industry is still in its early stages, the committee supported a 'wait-and-see' approach to government regulation. For their part, regulators can support this vision by ensuring equitable treatment under the law, low transaction frictions and low barriers to entry (Vong 2014).

# 7  Future Research

Despite our own research and experience in the field, there is a need to test empirically the propositions presented in this paper. As a nascent technology any claims, opinions and assumptions about the future of crypto-currencies require further in-depth analysis of the roles that the various stakeholders in the digital currency ecosystem are playing. Current studies being undertaken on consumer and merchant usage and awareness should provide a clearer picture of where the markets for future growth in this space are likely to come from. While the arguments in this paper focus on the positive impacts, there is also a need to explore inhibiting factors to the more widespread use of the technology, particularly with regards issues related to ease of use, security and its use for nefarious activities like money laundering and the 'darknet' marketplace.

# 8  Conclusion

The digital currency implementation in Australia will have effects on the three sectors, that is, payments, retail and banking. But retailers should be the most amenable to the introduction of digital currency use with sizable reduction of payment processing fees for payments made.

While digital currencies hold significant opportunities for innovation in Australia, a real concern is that the introduction and application of onerous regulations and taxation regimes could stifle and discourage investment in digital currencies in Australia and reduce the positive effects that could be obtained from stimulating the economy through digital currency usage. The ramifications of taxing bitcoins like a commodity most likely are still being felt and the ATO should consider a full examination of what affects this choice has had on the economy and innovation and consider taxing it appropriately like a normal financial payment. On the other hand, there is the view that a regulatory environment needs to be created that ensures the protection of consumers and preserves the integrity of the financial system and that of the tax base (Australian Bankers' Association Inc 2014).

To gain the first mover advantage it is imperative for Australia to develop an innovative digital currency technology sector, providing many thousands of high-value, knowledge-based jobs in the process. This will give Australia a chance to play a leading role in this dynamic new industry with universal ramifications. Expertise, products and services developed here could be deployed throughout the world giving the country the diversity it needs in the economy. In addition, Orban (2014) argues that the shift from an inflationary currency to a deflationary one brings with it an ecologically sustainable future shaped by the impact in energy use, resource allocation and management, and ways of living of the future